\documentclass[useAMS, referee]{mn2e}
\usepackage{graphicx}
\usepackage{amsmath}
\usepackage{hyperref}
\usepackage{amsfonts}

\title[Timing Studies on X Per and its QPO Feature]{Timing Studies on X-Per and Discovery of its Transient QPO Feature}

\author[Z. Acuner et al.]{Z. Acuner$^{1}$\thanks{E-mail: zeynepa@astroa.physics.metu.edu.tr(ZA); seyda@astroa.physics.metu.edu.tr (\c{S}\c{S}); inam@baskent.edu.tr (S\c{C}\.{I}); 
altan@astroa.physics.metu.edu.tr (AB); muhammed@astroa.physics.metu.edu.tr (MMS)}, S. \c{C}. \.{I}nam$^{2}$, \c{S}.\c{S}ahiner$^{1}$, M. M. Serim$^{1}$, A. Baykal$^{1}$, and J. Swank$^{3}$
\\ $^{1}$Physics Department, Middle East Technical University, 06531 Ankara, Turkey \\ $^{2}$Department of Electrical and Electronics Engineering, Ba\c{s}kent University, 06810 Ankara, Turkey \\ $^{3}$Astrophysics Science Division, Goddard Space Flight Center, NASA, Greenbelt, MD 20771, USA}

\begin{document}

\date{Received 2014}

\pagerange{\pageref{firstpage}--\pageref{lastpage}} \pubyear{2014}

\maketitle

\label{firstpage}

\begin{abstract}

We present timing analysis of {\emph{RXTE}}-PCA and {\emph{INTEGRAL}}-ISGRI observations of X Per between 1998 and 2010. All pulse arrival times obtained from the {\emph{RXTE}}-PCA observations are phase connected and a timing solution is obtained using these arrival times. We update the long-term pulse frequency history of the source by measuring its pulse frequencies using {\emph{RXTE}}-PCA and {\emph{INTEGRAL}}-ISGRI data. From the {\emph{RXTE}}-PCA data, the relation between frequency derivative and X-ray flux suggests accretion via the companion's stellar wind. On the other hand, detection of the transient QPO feature peaking at $\sim 0.2$ Hz suggests the existence of an accretion disc. We find that  double
break models fit the average power spectra well, which suggests that the source has at least two different accretion flow components dominating the overall flow.  From the power spectrum of frequency derivatives,  we measure a power law index of $\sim -1$ which implies that on short time scales disc accretion dominates over noise, while on time scales longer than the viscous time scales the noise dominates.  From pulse profiles, we find a correlation between pulse fraction and count rate of the source.    

\end{abstract}


\section{Introduction}

X Per (4U 0352+309) was revealed, using \emph{Ariel 5} and \emph{Copernicus} observations, to be a persistent low luminosity 
accreting pulsar with a pulse period of $\sim 836$ s (White et al. 1976). The neutron star in the binary system has a wide 
orbit around a Be type star with an orbital period of $\sim 250$ days and an eccentricity of $\sim 0.11$ 
(Delgado-Marti et al. 2001). 

The binary orbit of the system is wide and rather circular, such that the X-ray pulsar does not pass through the equatorial disc of the Be star. Therefore, X Per does not exhibit Type-I outbursts near periastron passages, despite the fact that such 
outbursts are typical of most Be X-ray binary systems.  Taking into account the observed X-ray luminosities ($L_x \sim 10^{35}$ erg s$^{-1}$) of the system, Delgado-Marti et al. (2001) suggested that accreting matter is 
supplied by a slow ($\sim$ 150 km s$^{-1}$) and dense wind possibly originating from the equatorial disc around the companion star.

Long term pulse period variation of the source has been monitored since 1970s using observations of various X-ray 
observatories (Nagase 1989; Lutovinov et al. 1994; Haberl 1994; Robba et al. 1996; Di Salvo et al. 1998; 
Delgado-Marti et al. 2001; Lutovinov et al. 2012). Before 1978, X Persei was spinning up with a rate of 
$\dot{P}/P \sim -1.5 \times 10^{-4}$ yr${^{-1}}$. Between 1978 and 2002 the source was in a long-term spin-down episode with
a rate of $\dot{P}/P \sim 1.3 \times 10^{-4}$ yr${^{-1}}$. After 2002, it was found that the spin rate changed sign again as 
the source was found to spin up with a rate of $\dot{P}/P \sim -3.6 \times 10^{-4}$ yr${^{-1}}$ which is significantly 
higher in magnitude compared to that of the previous spin-up episode before 1978. 

In this paper, we present results of the timing analysis of {\emph{RXTE}}-PCA and \emph{INTEGRAL}-ISGRI observations of X Per. In Section \ref{obssec}, observations are introduced. In Section \ref{timingsec}, our timing analysis is presented. 
In Section \ref{sumsec}, our results are summarized and discussed. 

\section{Observations}\label{obssec}

\subsection{{\emph{RXTE}}}\label{RXTEObs}

PCA (Proportional Counter Array) onboard {\emph{RXTE}} (\emph{Rossi X-ray Timing Explorer}) had five identical  co-aligned 
proportional counter units (PCUs) (Jahoda et al. 1996, 2006). Each PCU had an effective area of approximately 1300 $cm^{2}$. The PCA was able to detect photons having energies between 2 and 60 keV, with a spectral resolution of 18 per cent at 6 keV and a field of view (FOV) at full width at half-maximum (FWHM) of $\sim1\degr$. The number of active PCUs during the observations of X Per varied between 
one and three. Data from all the available PCUs are used in this analysis and count rates shown in Figure \ref{lcurve} are the values corrected as if 5 PCUs were operational.

X Per was observed by {\emph{RXTE}} between 1 July 1998 and 17 February 2003. The exposures of the individual pointed observations vary between 2 ks and 15 ks. The total exposure of 148 pointings adds up to $\sim793$ks (see Table \ref{RXTElist} for details). We extract light curves of the source with 0.1s time binning in  3 $-$ 20 keV energy band using \texttt{GoodXenon} mode events from PCA. 

\texttt{HEASOFT} v.6.13 software is used for the PCA data analysis. Only the data corresponding to times when the elevation angle is greater than 10$\degr$, the offset from the source is less than 0.02$\degr$ and the 
electron contamination of PCU2 is less than 0.1 are analysed. The latest background estimator models supplied by the {\emph{RXTE}} Guest Observer Facility are used  to extract background spectra and light curves. These background subtracted light curves are corrected to the barycenter of the Solar System as well as for binary motion using the orbital parameters of X Per (see Table 2 in Delgado-Marti et al. 2001). 

\begin{table}
\centering
\caption{{\emph{RXTE}}-PCA Observation List}
\label{RXTElist}
\begin{tabular}{ c | c | c | c }
\hline \hline
{\bf{Proposal ID}} & {\bf{Exposure}} & {\bf{Number of}} & {\bf{Time}} \\
		 & {\bf{(ks)}} & {\bf{Pointings}} & {\bf{(MJD)}} \\
\hline
30099 & 252 & 56 & 50995 $-$ 51406 \\
40424 & 60 & 14 & 51420 $-$ 51597 \\
50404 & 58 & 12 & 51634 $-$ 51940 \\
60067 & 398 & 58 & 52000 $-$ 52687 \\
60068 & 25 & 8 & 52094 $-$ 52398 \\
\hline \hline
\end{tabular}
\end{table}

\subsection{\emph{INTEGRAL}}\label{INTEGRALObs}

The Imager on-board the \emph{INTEGRAL} Satellite (IBIS) is a coded mask instrument optimized for high angular resolution (12$\arcmin$ at FWHM) (Ubertini et al. 2003). As a consequence of IBIS large FOV ($8\degr.3 \times 8\degr.0$ fully coded, $29\degr \times 29\degr$ zero response) and the dithering strategy of \emph{INTEGRAL} observations,  sky coverage is good and exposure times of detected sources reach up to millions of seconds. The IBIS consists of two detector layers operating in 
different energy bands. The data analyzed in this paper are obtained from the upper layer, \emph{INTEGRAL} Soft Gamma-Ray Imager (ISGRI), which operates in the energy range 15 keV -- 1 MeV, and has a time resolution of 61 $\mu$s. 

When the offset of the source of interest increases, the coding factor decreases resulting in increased uncertainties in the images, flux values and spectra. Therefore it is not recommended to include in the analysis observations in which the source lies outside the 50 per cent partially coded FOV ($19\degr \times 19\degr$).  Since X Per is a bright source, its flux is confidently determined even when its position is in the partially coded region. Consequently, the selection criteria for \emph{INTEGRAL} observations are; an off-axis angle of less than $10\degr$ and ISGRI good times of more than 1 ks. The publicly available 
pointing observations between MJD 53069 $-$ 55451 (5 March 2004 $-$ 12 September 2010) reach to a number of 766 science windows (SCWs), each having typical 
durations of 2-3 ks. These observations were analysed before by Lutovinov et al. (2012), but in this paper we present a reanalysis of these observations in order to measure pulse periods using a different technique described in Section \ref{ptsec}.

The latest version of standard data analysis software \verb"OSA v.10.0" is used for pipeline processing. Images in two energy bands (20 -- 40 and 40 -- 60 keV) are generated from IBIS-ISGRI data with an input catalogue consisting of strong sources in FOV: Crab, 3C 111, NGC 1275, IGR J02343+3229, XY Ari, 1H 0323+342, RX J0440.9+4431 and X Per. Background maps provided by the ISGRI team are used for background correction. 10 s binned light curves are generated by the tool \verb"II_LIGHT" and corrected to the Solar System barycenter. The effective exposure of the corrected IBIS-ISGRI light curve of X Per is around 2 Ms. Resulting light curves are also corrected for the binary motion using orbital parameters of X Per (see Delgado-Marti et al. 2001).

\section{Timing Analysis}\label{timingsec}

We use 0.1 s time binned {\emph{RXTE}}-PCA and 10 s time binned \emph{INTEGRAL} light curves as 
described in Sections \ref{RXTEObs} and \ref{INTEGRALObs}, for the timing analysis.

\begin{figure*}
 \begin{tabular}{l r}
 \includegraphics[width=7.3cm]{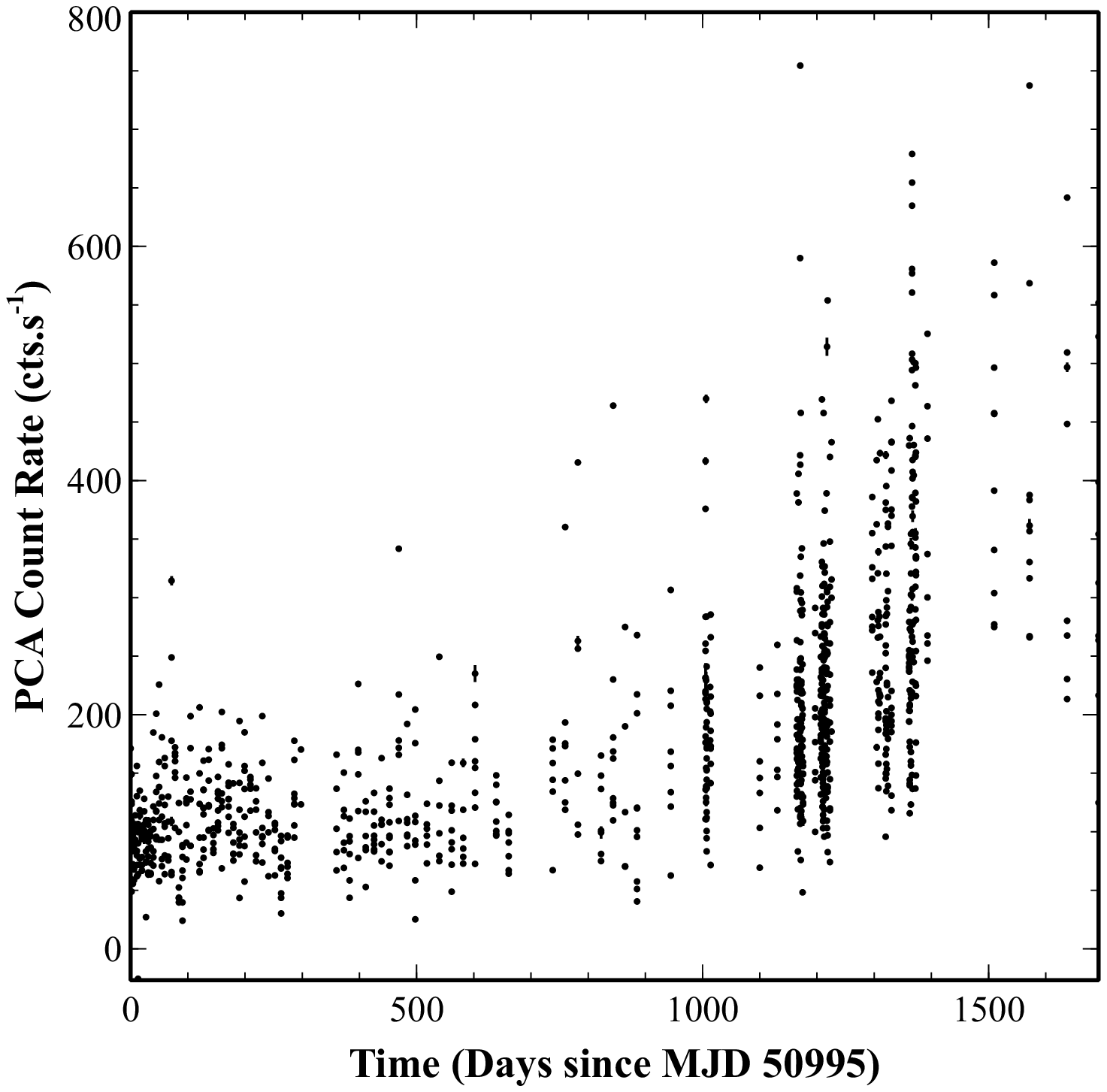} &
 \includegraphics[width=7.3cm]{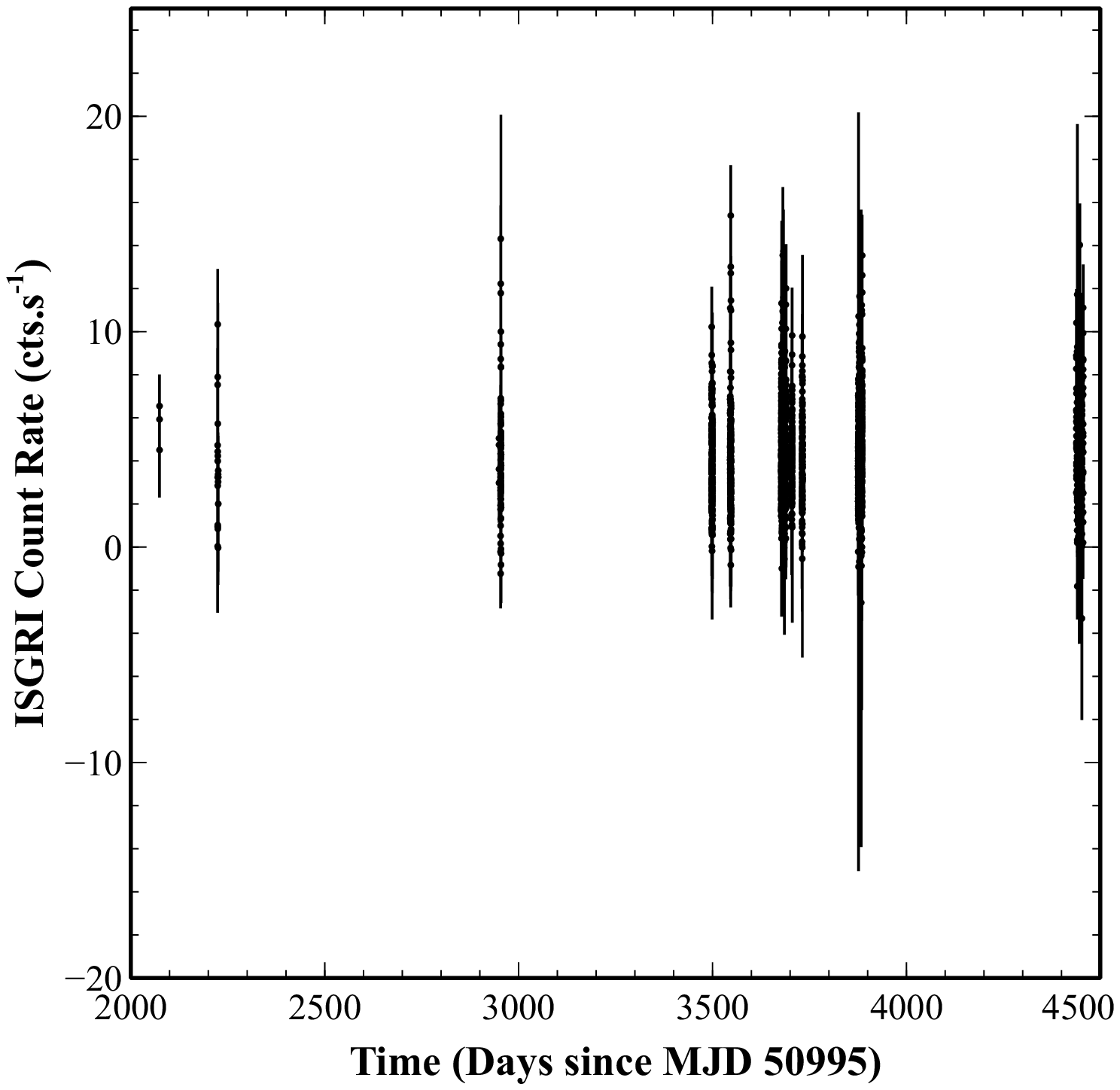} \\
 \end{tabular}
 \caption{837s binned light curves obtained from {\bf{(left)}} {\emph{RXTE}}-PCA (3 -- 20 keV, corrected for 5 PCUs) and 
 {\bf{(right)}} \emph{INTEGRAL} IBIS-ISGRI (20 -- 40 keV) observations.}
 \label{lcurve}
\end{figure*}

In Figure \ref{lcurve}, overall 837s binned {\emph{RXTE}}-PCA and \emph{INTEGRAL} light curves of the source are presented.

\subsection{Pulse Timing}\label{ptsec}

\begin{figure*}
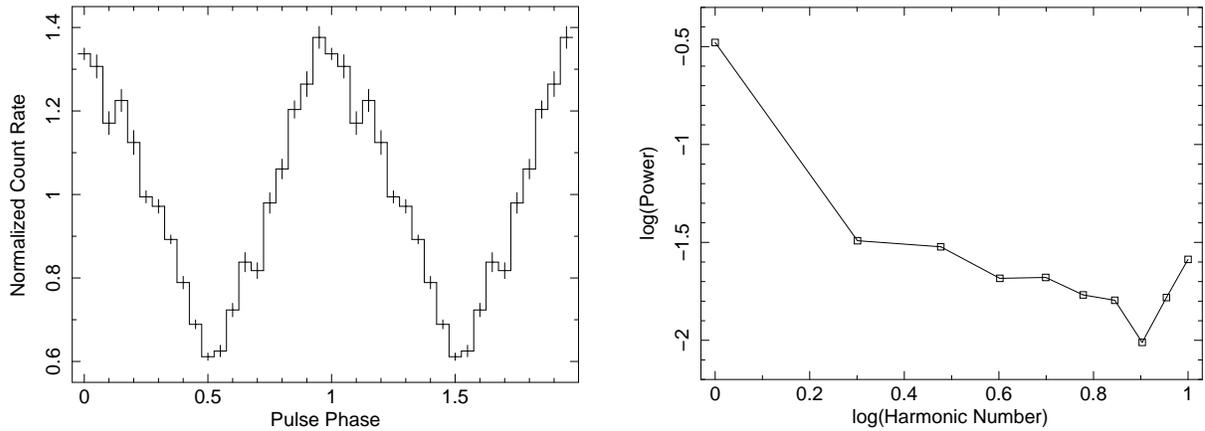

 \begin{tabular}{l r}
 \includegraphics[width=5.6cm, angle=-90]{pulseprofile.ps} &
 \includegraphics[width=5.6cm, angle=-90]{powerpulse.ps} \\
 \end{tabular}
 \caption{A typical pulse profile (left) obtained from {\emph{RXTE}}-PCA and its power spectrum (right) in terms of harmonic number.}
 \label{pulseprof}
\end{figure*}

In order to measure pulse periods of X Per, we fold time series on statistically independent trial periods (Leahy et al. 1983). We construct template pulse profiles by folding the data on the period corresponding to maximum $\chi^2$. Pulse profiles consisting of 20 phase bins are represented by their Fourier harmonics (Deeter \& Boynton 1985). In Figure \ref{pulseprof}, a typical pulse profile and the corresponding power spectrum in harmonic number are given.

\begin{figure*}
 \center{\includegraphics[width=12cm]{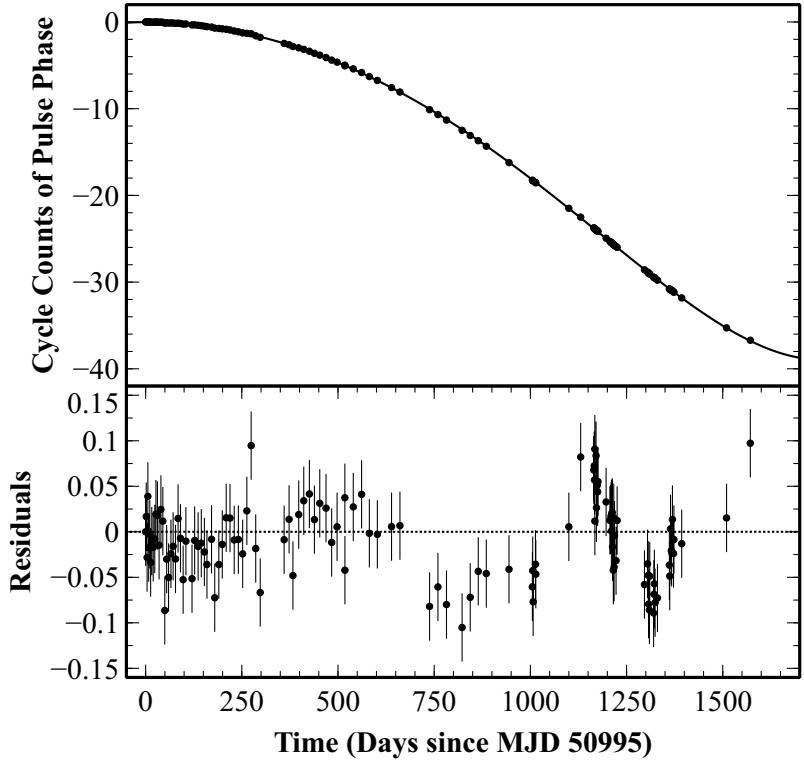}}
 \vspace{-2cm}
 \caption{Pulse arrival times and their residuals after a fifth order polynomial fit obtained from {\emph{RXTE}}-PCA 
 observations.}
 \label{arrt}
\end{figure*}

We have connected in phase all pulse arrival times of RXTE observations over 1550 days time span. In the phase connection process, we construct pulse arrival times
for a time span where the maximum phase shift is less than 1. In this way, we avoid cycle count ambiguity.
This time scale for X Per is around 220 days. We divide the total time span into 8 time intervals, each around 220 days. Then we measure pulse arrival times with respect to the best period corresponding to that time interval. Then we align the slopes of the pulse arrival times in overlapping time intervals. The pulse arrival times thus measured are presented in the upper panel of Figure \ref{arrt}.

In order to see the effect of pulse shape fluctuations, we estimate pulse arrival times using first, second, fifth and tenth harmonic numbers. Analyzing each set in the same way, we obtain residuals consistent with each other within 1$\sigma$ level. Therefore we conclude that pulse shape variations do not cause drastic changes in the pulse timing analysis.

The phase connected pulse arrival times in Figure \ref{arrt} are fitted to the fifth order polynomial,
\begin{equation}
\delta \phi = \delta \phi_{o} + \delta \nu (t-t_{o})
+ \sum _{n=2}^{5} \frac{1}{n!}
 \frac {d^{n} \phi}{dt^{n}} (t-t_{o})^{n}
\label{polyn}
\end{equation}
where $\delta \phi$ is the pulse phase offset obtained from pulse timing analysis, $t_{o}$ is the mid-time of the observation; 
$\delta \phi_{o}$ is the residual phase offset at t$_{o}$; $\delta \nu$ is the correction of pulse frequency at time $t_0$;
$\frac {d^{n} \phi}{dt^{n}} $ for $n$=2,3,4,5 are first, second, third and fourth order derivatives of pulse frequency.

With only the statistical errors on the arrival phases $\chi^2$ per degree of freedom, being 9.5, is not acceptable. Systematic errors or short term variations not captured in the fifth order polynomial could be responsible. The errors have been multiplied by a factor of 3 such that the reduced $\chi^2$ is 1, in order to obtain  errors on the polynomial fits to the longer term variations. Pulse arrival times (pulse cycles) and the residuals of the fit after removal of the fifth order polynomial trend are shown in Figure \ref{arrt}. In Table \ref{timingtable}, we present the
timing solutions for  X Per between MJD 50995 and 52562.

\begin{table}
\centering
\caption{Timing Solution of X Per between MJD 50995 and 52562.}
\label{timingtable}
\begin{tabular}{l | r}
\hline \hline
{\bf{Parameter}} & {\bf{Value}} \\
\hline
Epoch ($t_0$) (Days in MJD) & 50995.038(1) \\
{\bf{Timing parameters at $t_0$:}} & \\
~~Spin Period (s) & 837.666(6) \\
~~Spin Frequency (Hz) & 1.19379(9)$\times 10^{-3}$ \\
~~$\dot{\nu}$ (Hz s$^{-1}$) & -5.5(3)$\times 10^{-15}$ \\
~~$\ddot{\nu}$ (Hz s$^{-2}$) & 6.6(6)$\times 10^{-23}$ \\
~~$\dddot{\nu}$ (Hz s$^{-3}$) & -3.6(9)$\times 10^{-30}$ \\
~~$\ddddot{\nu}$ (Hz s$^{-4}$) & 9(2)$\times 10^{-38}$ \\
\hline \hline
\end{tabular}
\end{table}

\begin{figure*}
\begin{tabular}{c}
 \includegraphics[width=12cm]{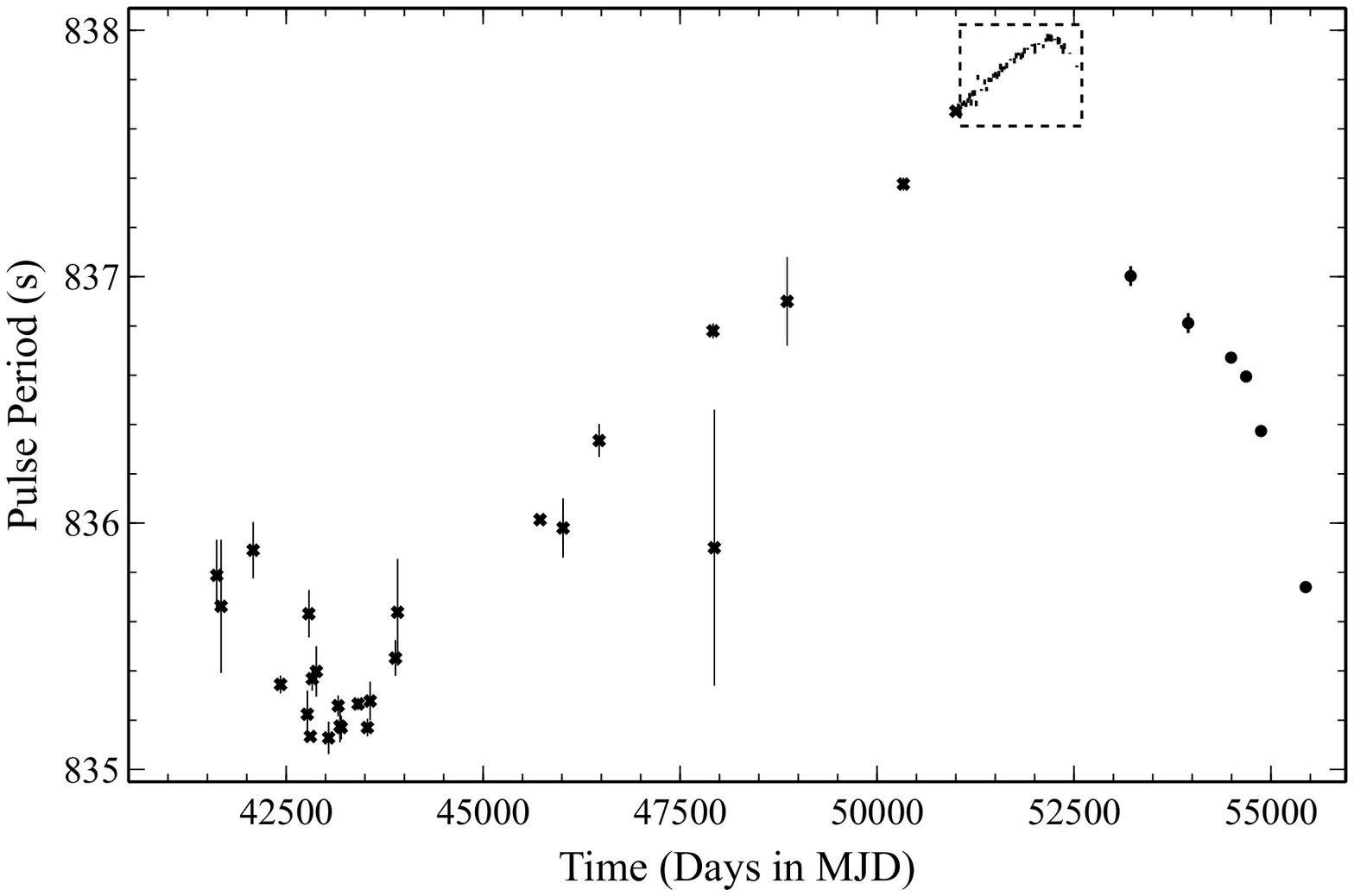} \\
 \includegraphics[width=12cm]{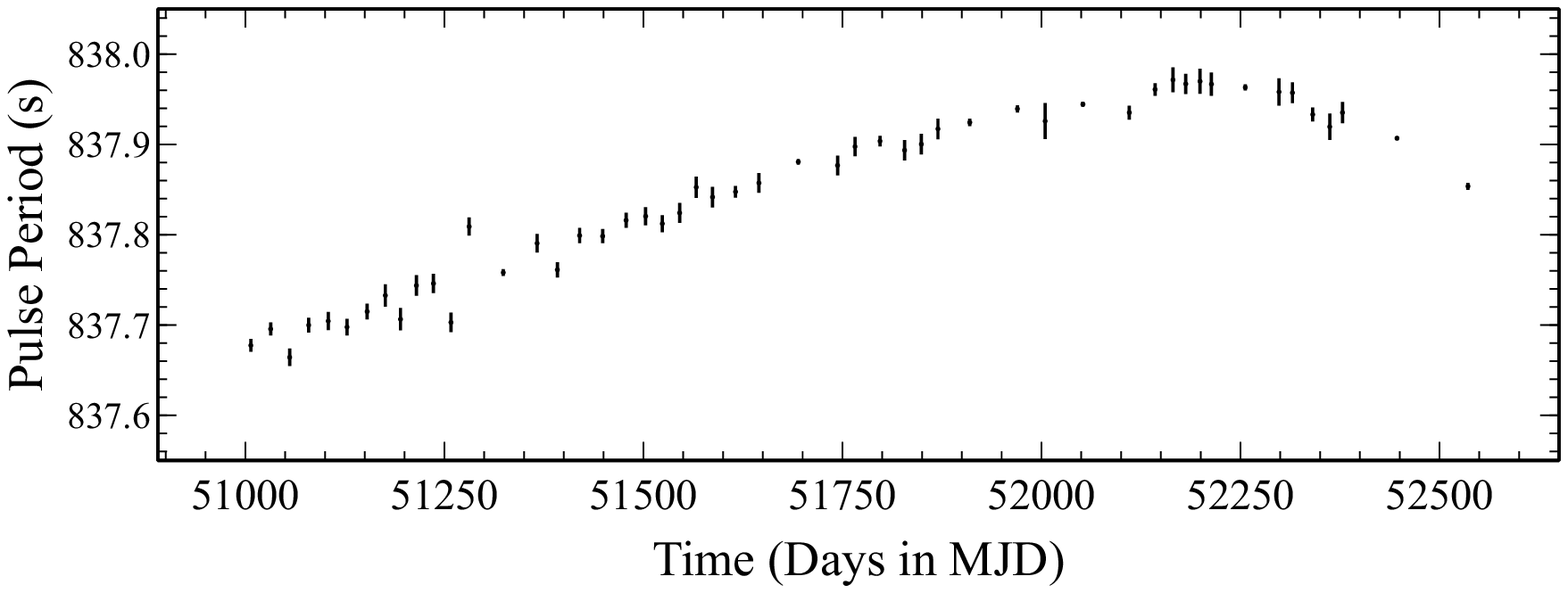} \\
 \end{tabular} 
 \caption{{\bf{(top)}} Pulse period history of X Per. Pulse periods indicated with cross sign are previous measurements 
 obtained from various observatories (see references in Lutovinov et al. 2012). Filled circles represent 
 \emph{INTEGRAL} measurements in this work. Other data points enclosed in the dashed rectangle are {\emph{RXTE}}-PCA measurements of this work. 
 {\bf{(bottom)}} Closer view of the pulse period evolution obtained from {\emph{RXTE}}-PCA observations.}
 \label{history}
\end{figure*}

In order to measure pulse frequencies, we fit a linear model to arrival times from $\sim 20 - 30$ days intervals. The slopes of these linear fits give us estimates of the pulse frequency values at mid-times of corresponding intervals. Resulting pulse periods of the source are presented together with the previous pulse period measurements of the source (see references in Lutovinov et al. 2012) in Figure \ref{history}.  In Figure \ref{history}, we do not include measurements from {\emph{RXTE}} and {\emph{INTEGRAL}} by Lutovinov et al. (2012), since the values of their measurements are not reported numerically, but when we compare the scanned values in Figure 2 of Lutovinov et al. (2012), it is seen that our measurements are in agreement with theirs. In Table \ref{pertab}, we present our pulse period measurements.

\begin{table}
 \caption{Pulse periods of X Per measured in this work. Numbers in parantheses indicate the uncertainties in the least significant figures.}
 \label{periods}
 \center{\begin{tabular}{cccc}
 \hline
 \hline	
\bf{Epoch} & \bf{Pulse Period} 	& \bf{Epoch} & \bf{Pulse Period} \\
\bf{(MJD)} & \bf{(s)} 		& \bf{(MJD)} & \bf{(s)}		 \\
 \hline
 51006.9 &	 837.678(7) & 51797.3 & 837.904(6) \\
 51031.8 &	 837.696(7) & 51828.1 & 837.89(1)  \\
 51055.7 &	 837.66(1)  & 51849.1 & 837.90(1)  \\
 51079.5 &	 837.700(8) & 51870.0 & 837.92(1)  \\
 51104.1 &	 837.70(1)  & 51909.9 & 837.924(4) \\
 51127.7 &	 837.698(9) & 51969.8 & 837.939(4)  \\
 51153.0 &	 837.715(9) & 52004.7 & 837.93(2)  \\
 51175.8 &	 837.73(1) & 52052.1 & 837.944(3) \\
 51194.9 &	 837.71(1) & 52110.4 & 837.935(8) \\
 51214.8 &	 837.74(1) & 52142.7 & 837.961(7) \\
 51236.3 &	 837.74(1) & 52165.2 & 837.97(1) \\
 51258.3 &	 837.70(1) & 52181.3 & 837.97(1) \\
 51281.1 &	 837.81(1) & 52199.2 & 837.97(1) \\
 51323.9 &	 837.758(4) & 52213.5 & 837.97(1) \\
 51366.4 &	 837.79(1) & 52256.0 & 837.963(3) \\
 51392.0 &	 837.761(8) & 52298.4 & 837.96(2) \\
 51419.9 &	 837.799(9) & 52315.3 & 837.96(1) \\
 51448.8 &	 837.798(8) & 52340.8 & 837.933(8) \\
 51478.3 &	 837.816(8) & 52362.1 & 837.92(1) \\
 51502.8 &	 837.82(1) & 52378.2 & 837.94(1) \\
 51523.8 &	 837.812(9) & 52446.6 & 837.907(2) \\
 51545.5 &	 837.82(1) & 52535.8 & 837.854(4) \\ 
 51566.4 &	 837.85(1) & 53219.2$^a$ & 837.00(4) \\
 51586.9 &	 837.84(1) & 53949.3$^a$ & 836.81(4) \\
 51615.7 &	 837.847(6) & 54494.0$^a$ & 836.67(2) \\
 51645.1 &	 837.86(1) & 54684.5$^a$ & 836.595(4) \\
 51694.5 &	 837.881(3) & 54873.8$^a$ & 836.37(2) \\
 51743.9 &	 837.88(1) & 55441.1$^a$ & 835.7402(3) \\
 51765.9 &	 837.90(1) &	 &			 \\
 \hline
 \hline
\end{tabular}} \\
\begin{flushleft}
$^{a}$ Pulse periods are measured from \emph{INTEGRAL} observations.
\end{flushleft}
\label{pertab}	
\end{table} 

For timing analysis of \emph{INTEGRAL} observations, we use 10s binned background corrected 20 -- 40 keV IBIS-ISGRI  light curves of the source. From the light curves having $\sim7-10$ days time spans, we measure the best pulse frequency  by folding the light curve on statistically independent pulse frequencies. Then we estimate the pulse arrival times by constructing master and sample pulses as described above for the {\emph{RXTE}} analysis. From the slopes of the pulse arrival times 
($\delta \phi = \delta \nu (t-t_{o})$) we obtain corrections to the pulse frequencies. In Figure \ref{history} and Table \ref{pertab}, we present pulse periods obtained from \emph{INTEGRAL} observations.

\begin{figure*}
 \begin{tabular}{l r}
 \includegraphics[width=7.6cm]{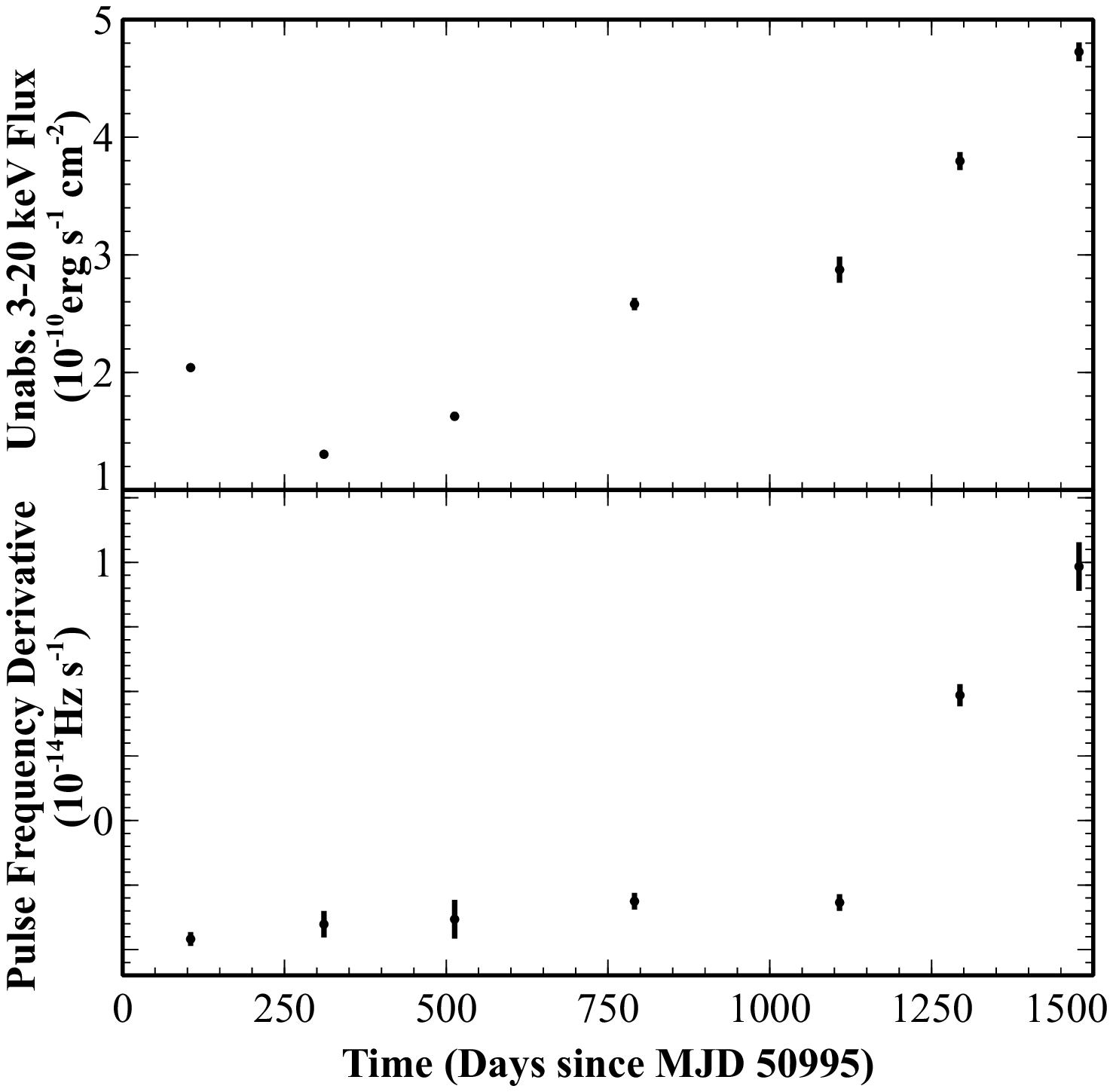} &
 \includegraphics[width=7.4cm]{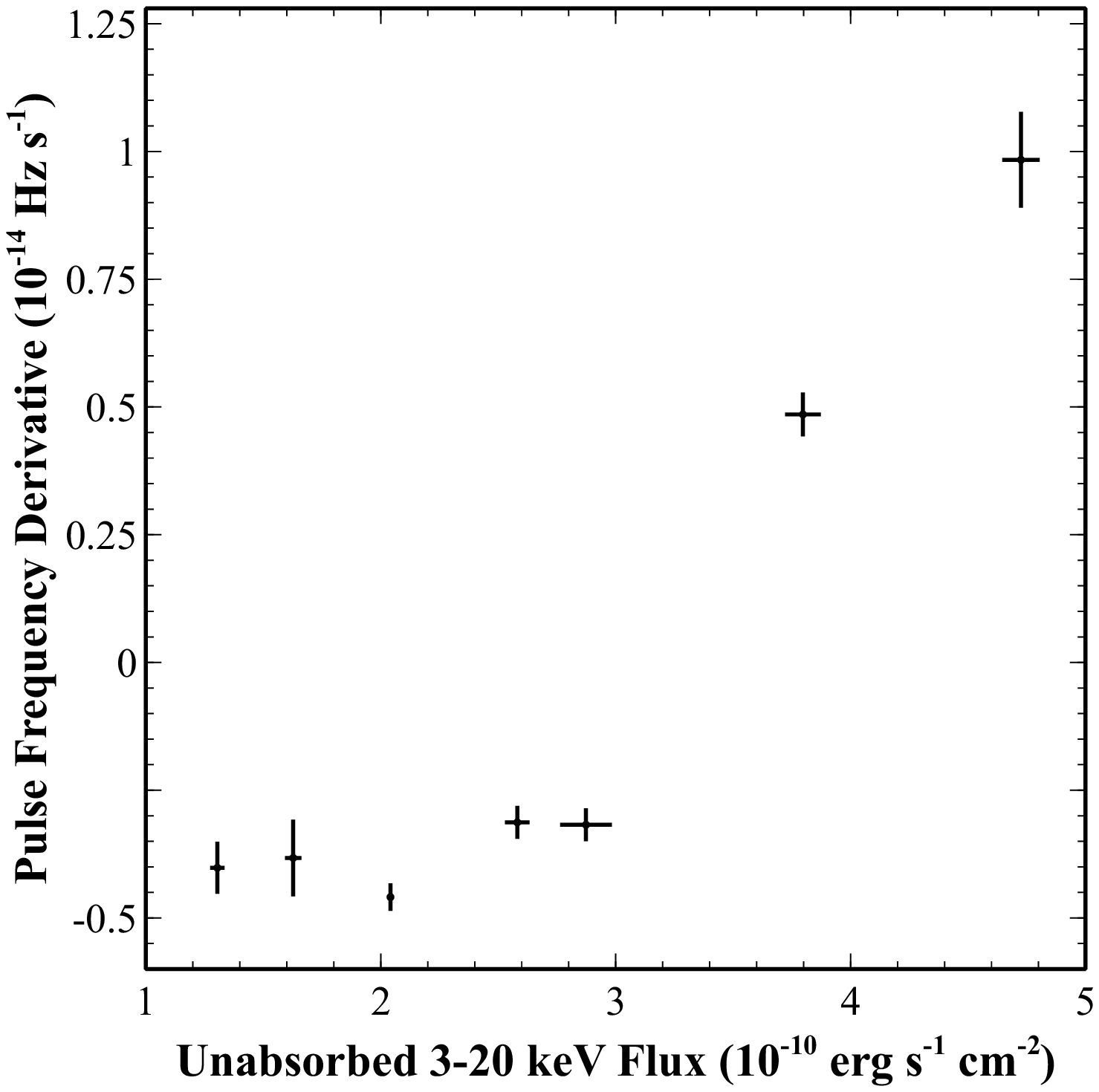} \\ \end{tabular}
 \caption{{\bf{(left)}} Unabsorbed 3 -- 20 keV {\emph{RXTE}}-PCA flux and frequency derivative as a function of time, {\bf{(right)}} frequency derivative as a function unabsorbed 3 -- 20 keV \emph{RXTE}-PCA flux.}
 \label{fluxfreqtime}
\end{figure*}

We measure pulse frequency derivatives by fitting quadratic polynomials to $\sim 200$ day long stretches of pulse arrival 
times from {\emph{RXTE}}-PCA observations. The quadratic coefficients of the fits give the pulse frequency 
derivatives. Corresponding 3 -- 20 keV X-ray flux values are calculated by modelling each corresponding spectrum with an absorbed black body and power law model, a model previously used for this source (Lutovinov et al. 2012). In Figure \ref{fluxfreqtime}, we present frequency derivatives and corresponding flux measurements as a function of time and frequency derivative values as a function of unabsorbed 3 -- 20 keV X-ray flux.

\subsection{Broad Excess and QPO Features}

Intensity power spectra obtained using
\texttt{powspec} of \texttt{HEASOFT} from 0.3s binned light curves 
of each {\emph{RXTE}}-PCA observation are used to look for further periodicities other than those caused by spin and orbital modulations. The power spectra usually exhibit a broad excess around $\sim$ 0.01 -- 0.08 Hz and occasionally exhibit a narrow excess peaking at $\sim 0.2$ Hz that  can be interpreted as a transient quasi-periodic oscillation (QPO) feature.

In order to analyze broad excess and the possible QPO feature, all power spectra are fitted with a double power law. For the power spectra with
a broad excess, a broad Gaussian is added to the model. Where a narrow excess is present, a Lorentzian is added to the model.

The broad excess feature appears between 0.01 -- 0.02 Hz at the earlier observations and gradually shifts towards higher
frequencies in time. This excess below 0.1 Hz can be observed in
almost 90 per cent of the power spectra produced and it  is mainly prominent
between 0.02 Hz and 0.08 Hz. Takeshima (1998) found a 0.054 Hz QPO feature before, in 1996 \emph{RXTE} data, which lies in this frequency range. In our analysis of
the later \emph{RXTE} data, we find the activity in this range only as broad excess features, which are strong with high RMS amplitudes yet with coherence values too low to be considered QPOs.

In contrast, at around 0.2 Hz, we occasionally
observe narrow peaks which can be
interpreted as QPOs. For each observation, in order to test the significance of these QPOs,
we average 4 power spectra and rebin the frequencies by a factor 4. Each resulting power spectrum is then normalized by dividing it by the continuum model consisting of a double power law and a broad Gaussian if a broad excess is present and multiplied by 2 (van der Klis
1989). The final power spectrum would be consistent with a chi square
distribution for a degree of freedom $2\times 4\times 4$ = 32.  
According to this formalism, an excess power of 7 corresponds to a total power of $7\times 4\times 4$ = 112. The probability of detecting a false
 signal becomes ${\rm{Q}}(112|32)=8.2\times 10^{-11}$. Since we have 128 independent
 frequencies in each  power
spectra, the total probability of having a false signal becomes $ 128
\times 8.2\times 10^{-11} = 1.0\times 10^{-8}$ which corresponds to
a more than $5.73\sigma$ level of detection. 

\begin{figure*}
\begin{tabular}{l r}
\includegraphics[bb=580 0 109 692,  angle=-90,
width=7.3cm]{110604f4r.ps} &
\includegraphics[width=6.9cm]{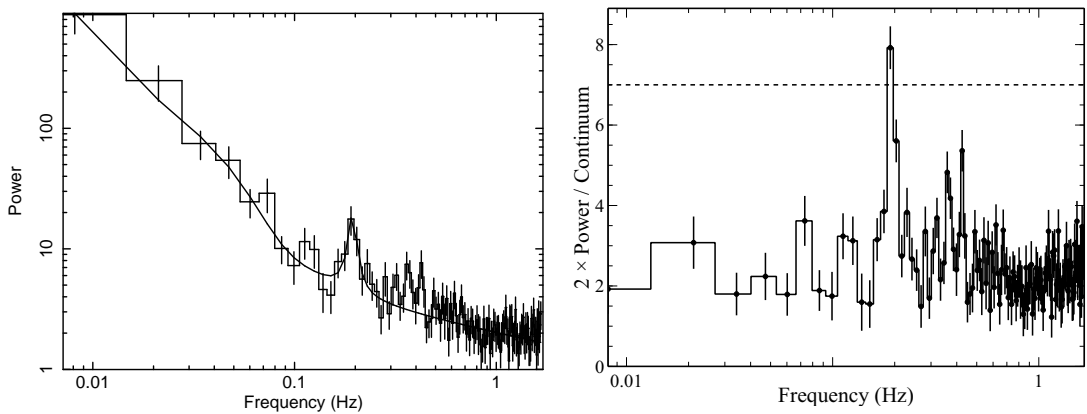}
\\ \end{tabular}
\caption{{\bf{(left)}} Power spectrum rebinned by a factor of 4 obtained from the {\emph{RXTE}}-PCA observation ID 30099-01-19-00 (at
MJD 51060). QPO feature modelled as a Lorentzian peaked at 0.195 Hz is evident. {\bf{(right)}} Power spectrum rebinned by a factor of 4, multiplied by 2 and divided by the continuum modelled with a double power law with the addition of a broad Gaussian. Using the method of van der Klis (1989), power value of 7 corresponds to a false signal
probability of $1\times 10^{-8}$ which is indicated as a horizontal
dashed line.}
 \label{qpo}
\end{figure*}

We find that only one of the QPOs peaking at $\sim 0.2$ Hz has a significance at this level, the one at MJD 51060 (observation ID 30099-01-19-00). In Figure \ref{qpo} we present the intensity power
spectrum and the power spectrum of this observation,
 normalized with respect to the  continuum model. 
In the right panel of this figure, the $5.73\sigma$ level of detection is indicated as a horizontal dashed line, and it is seen that there is only one feature above this level. For this observation, the continuum is modelled with a double power
 law model with indices $-0.36\pm 0.09$ and $-2.19\pm 0.30$
  with the addition of a Gaussian peaking at 0.016 Hz to account for the broad excess. The QPO is modelled as a 
Lorentzian peaked at 0.195 Hz with a width of 0.018 Hz. The QPO 
has a coherence (i.e. the peak frequency divided by the width)  of 10.8 with an RMS amplitude of 4.3\%,
 whereas the broad power excess has a coherence of 0.77 with an RMS amplitude of 8.5\%.  
 
 In order to see the effect of 
time binning on power spectra, we rebin the lightcurve in 0.125s and 
0.25s and reconstruct the power spectra. In all cases, QPO detections are found to have similar significances.

\subsection{Average Power Spectra}

\begin{figure*}
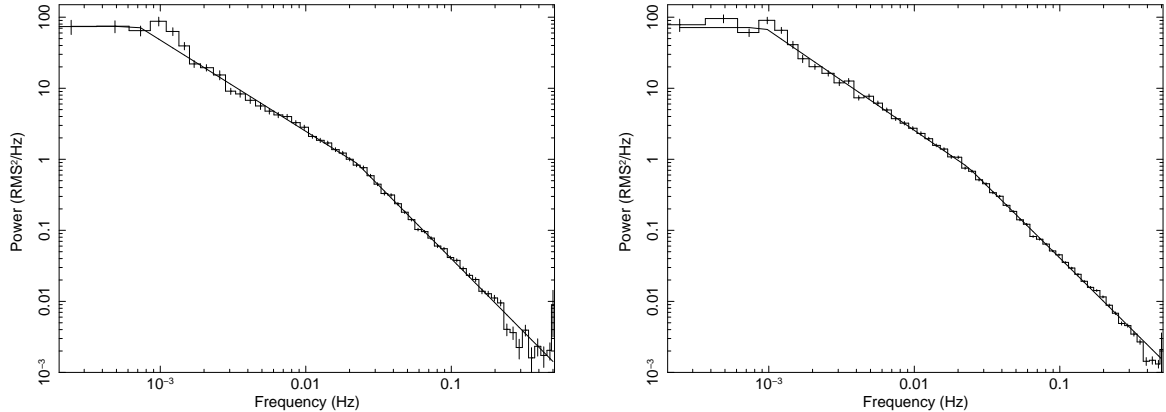

 \begin{tabular}{c c}
 \includegraphics[width=5.5cm,angle=-90]{30099_2.ps} &
 \includegraphics[width=5.5cm,angle=-90]{60067_2.ps} \\
 \end{tabular}
 \caption{Average (summed) power spectra obtained from the {\emph{RXTE}}-PCA observations {\bf{(left)}} with proposal ID 30099 and 
 {\bf{(right)}} with proposal ID 60067. For each spectrum, the value of the bin which corresponds to pulse frequency is removed by substituting it by the average of two neighbouring bins.}
 \label{ltpow}
\end{figure*}

Two power spectra are produced to examine the long term averaged power spectral properties of X Per. Light curves obtained from observations with proposal ID 30099 (between MJD 50995 and MJD 51406) and a part of observations with proposal ID 60067 (between MJD 52208 and MJD 52687) are used to produce these two white noise subtracted power spectra using \texttt{powspec} of \texttt{HEASOFT} (see Figure \ref{ltpow}). Each average power spectrum is calculated by averaging the power spectra obtained from  \emph{RXTE} lightcurve portions in 1 s bins, each with a time
span of 4096s. We avoid choosing larger time intervals due to scarcity of data at the end of the {\emph{RXTE}} observation time span. It is convenient to pick out smaller time segments since they demonstrate the spectral behaviour of their broader counterparts. For both power spectra,  the
bin corresponding to the pulse frequency is replaced by the average of two neighbouring bins.    

We find that a double break power law model fits well to these power spectra. In this model, there are two breaks: The first break separates regions with $f^0$ (constant) and $f^{-1.4}$ dependence and the second break separates regions with $f^{-1.4}$ and $f^{-2}$ dependence. We find for the first break frequencies as $7(2) \times 10^{-4}$ Hz and $9.3(8) \times 10^{-4}$ Hz, respectively for the first and second power spectra. For the second break frequency we find
$2.3(2) \times 10^{-2}$ Hz and $2.36(9) \times 10^{-2}$ Hz for the two power spectra.  Although the power spectra are found to be similar, the average unabsorbed 3 -- 20 keV  X-ray luminosities obtained from the energy spectra vary significantly, from $\sim2.4\times 10^{34}$ erg s$^{-1}$ to $7 \times 10^{34}$ erg s$^{-1}$ for the first and second power spectra respectively (assuming a source distance of 950 pc). 

\subsection{Power Spectra of Pulse Frequency Derivatives}\label{tnoise}

 In order to compare torque fluctuations of X Per with
 other accreting X-ray pulsars, the power density spectrum for the pulse frequency variations is constructed. The red noise power density and associated random walk noise
 strengths are obtained using the technique developed by Deeter and Boynton (1982) and
 Deeter (1984). Some important properties are summarized here. For the $r$th-order red noise with strength S$_{r}$, mean square residual for the data spanning an interval
 with length T is proportional to S$_{r}$T$^{2r-1}$.

Expected mean square residual, after removing a polynomial of degree m over an interval of length T, is given by

\begin{equation} 
<\sigma _{R}^{2}(m,T)> = S_{r}T^{2r-1}<\sigma _{R}^{2}(m,1)>_{u},
\end{equation}
where $<\sigma _{R}^{2}(m,1)>_{u}$  
is the proportionality
factor which can be estimated by measuring the variance of residuals by removing the degree of polynomial $m$ for unit noise strength 
($S_1$). We estimate this factor by simulating time series for 
 $r$th-order red noise process for a unit noise strength 
($S_1$) for X Per observations. Our expected proportionality factors are consistent with those obtained by direct mathematical evaluation (Deeter 1984, see also Cordes 1980).

We obtain noise strengths at lower frequencies as f=1/T$_{max}$,
 where T$_{max}$ is the maximum time span of the pulse frequency history,
from the residuals of pulse frequencies by removing their linear trends.
For the higher frequencies, we have 
f$_{n}$=n/T$_{max}$, where n is a positive integer and we remove quadratic
trends from the pulse arrival times obtained from {\emph{RXTE}} data.

In order to check whether noise strengths are
stable or not,  we estimate an alternative power spectra by removing quadratic trends in pulse frequency at longer time scales and cubic trends in pulse arrival times at shorter time intervals.

\begin{figure*}
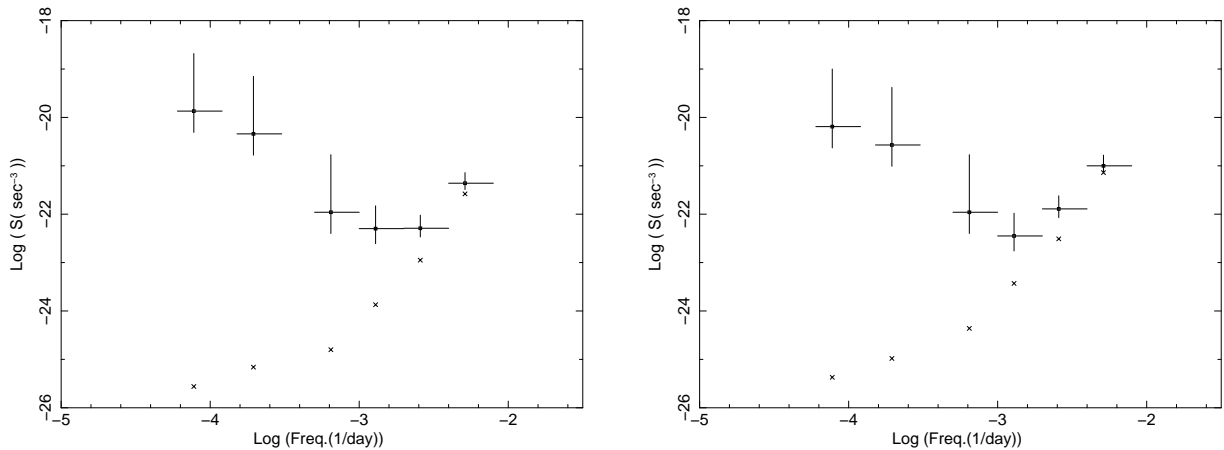

 \begin{tabular}{c c}
 \includegraphics[width=6cm,angle=-90]{noise_quad.ps} &
 \includegraphics[width=6cm,angle=-90]{noise_cubic.ps} \\
 \end{tabular}
 \caption{Power spectra estimated with respect to frequency obtained by removing {\bf{(left)}} quadratic trends and {\bf{(right)}} cubic trends. The crosses represent the power resulting from measurement noise.}
 \label{fnoise}
\end{figure*}

In Figure \ref{fnoise}, we present power
spectra estimates (or noise strengths) with respect to  
frequency f=n/T (or reciprocal of the time scale).
We find that both power spectra are consistent with each other
in terms of the average noise strength
$S_{r}$ and slope of the power spectra -0.85 -- -1.47 with a noise strength
$10^{-20}-10^{-23}$ Hz s$^{-2}$ for frequencies between 1/35 yr$^{-1}$ 
and 1 yr$^{-1}$.

\subsection{Pulsed Fraction}

\begin{figure}
 \centering
 \includegraphics[width=0.45\textwidth]{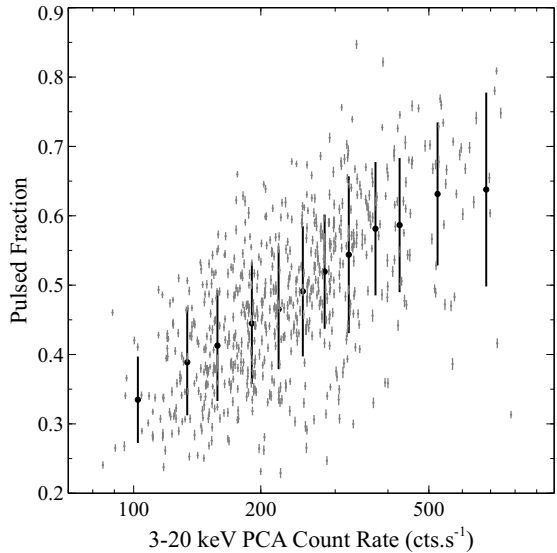}
 \caption{Variation of the pulsed fraction with mean count rate of pulsations. Grey data points are the unbinned results, whereas 
 the black data points are the results rebinned according to count rates.}
 \label{pf}
\end{figure}

We investigate the pulsed fraction variation of X Per for all the available {\emph{RXTE}}-PCA data between MJD 50995 and 52687 in 3$-$20 keV energy band. Using the timing solution stated in Table \ref{timingtable}, we extract  671
individual pulses of the source. Then we calculate the pulsed fraction and the mean count rate of 
each pulsation. The pulsed fractions are calculated with the standard definition;

\begin{equation}
 {\rm{Pulsed\,\,Fraction}}=\frac{p_{\mathrm{max}}-p_{\mathrm{min}}}{p_{\mathrm{max}}+p_{\mathrm{min}}}
\end{equation}

where $p_{\mathrm{min}}$ and $p_{\mathrm{max}}$ refers to minimum and maximum counts of the pulse. Pulsed fractions are then rebinned according to the  mean count rate of the pulsations. The results indicate that the pulsed fraction of the source correlates with the mean count rate (see Figure \ref{pf}).

\section{Discussion}\label{sumsec}

In this paper, we present our results of timing analysis of {\emph{RXTE}} and \emph{INTEGRAL} observations of X Per. Firstly, using the cross correlation technique, we add new measurements to the pulse period history presented by Lutovinov et al. (2012) (see Table \ref{periods}). 

From the right panel of Figure \ref{fluxfreqtime}, it is seen that frequency derivative of the source correlates with X-ray flux when the source spins up or when the source spins down and the 3-20 keV X-ray flux is above $\sim 3\times 10^{-10}$erg s$^{-1}$cm$^{-2}$, while frequency derivative does not vary significantly when source spins down and 3 -- 20 keV X-ray flux is below $\sim 3\times 10^{-10}$erg s$^{-1}$cm$^{-2}$. Figure \ref{fluxfreqtime} resembles Figure 1 of Shakura et al. (2012) which presents frequency derivative as a function of normalized mass accretion rate (which should be positively correlated with the X-ray flux) of a neutron star accreting from stellar wind. In Figure 1 of  Shakura et al. (2012), a minimum spin-down rate value is reached when mass accretion rate is equal to a critical mass accretion rate ($\dot{M}_{\mathrm{cr}}$) above which spin rate is correlated with mass accretion rate. Mass accretion rates below or above this value lead to smaller spin-down rates and above a certain normalized mass accretion rate value ($y_0$), the source enters a spin-up regime. This similarity may be an indication of the fact that the primary source of accreted matter is stellar quasi-spherical stellar wind of the companion.  However, the transient QPO feature of the source suggests a transient formation of an accretion disc as the neutron star accretes via the companion's stellar wind. Another low-luminosity long-orbital-period persistent accretion powered pulsar in a Be -- X-ray binary system, 1RXS J225352.8+624354, has recently been found to show signs of a possible transient accretion disk formation (Esposito et al. 2013).

Quasi Periodic Oscillations (QPOs) with peak frequencies between 0.01 and 0.4 Hz  have been observed in many accretion powered pulsars (see \.{I}nam et al. 2004 and references therein). The QPO peak frequencies ($\sim 0.1-0.2$ Hz) and RMS amplitudes ($\sim 4\%$) of X Per are typical. 

Using a Keplerian frequency model (van der Klis et al. 1987), QPO frequency is interpreted as the Keplerian frequency at the inner radius of the accretion disc and can be expressed as
\begin{equation}
\nu_k = {{1} \over {2\pi}}\left(\frac{GM}{r_0^3}\right)^{1/2}
\end{equation}
where $M$ is the neutron star mass which is taken as $1.4 M_{\odot}$, $r_0$ is the radius of the inner disc and $G$ is the universal gravitational constant.  Using Keplerian frequency model, we estimate inner disc radius ($r_0$) as $\sim 4.99 \times 10^{8}$ cm for the most prominent QPO observed at MJD 51060 with a frequency of 0.192 Hz. It is important to note that as the X Per's spin frequency is too small compared to its QPO frequency, it is not observationally possible to discriminate between Keplerian frequency model and beat frequency model (Alpar and Shaham 1985).

Inner disc radius can also be expressed as (Ghosh and Lamb 1979),
\begin{equation}
r_0 \sim 0.52 \mu^{4/7} (2GM)^{-1/7} \dot{M}^{-2/7}
\end{equation}
 where $\mu = B \times R^3$ is the magnetic dipole moment, $R$ is the neutron star radius, taken as $10^6$ cm, $\dot{M}$ is the mass accretion rate. At MJD 51060, when the most prominent QPO is observed, 3-20 keV luminosity of the source is $4.38\times 10^{34}$ erg s$^{-1}$. This corresponds to a mass accretion rate of $\sim 2.4 \times 10^{14}$ g s$^{-1}$ assuming that this luminosity value is of the order of the bolometric luminosity and using the relation $L\simeq GM\dot{M}/R$. Then, the value of $3.56 \times 10^{11}$ Gauss is estimated for the surface dipole magnetic field of the source. This magnetic field estimate is roughly comparable to some previous magnetic field estimates ($\sim 2.5 \times 10^{12}$ Gauss) of Lutovinov et al. (2012) and Coburn et al. (2001) based on cyclotron resonance scattering feature in the energy spectra, but our estimate is not consistent with the magnetar-like magnetic fields proposed for this source by Doroshenko et al. (2012).
 
Power spectra of pulse frequency derivatives have been studied for several accretion powered X-ray pulsars. Random walk in pulse frequency or white noise in pulse frequency derivative are suitable models for wind accreting systems
like Vela X-1, 4U 1538-52 and GX 301-2 (Deeter 1981,
Deeter et al. 1989, Bildsten et al. 1997). They have flat power spectra with white noise strengths in the range $10^{-20}-10^{-18}$ Hz s$^{-2}$. For 4U 1907+09 random walk in pulse frequency is the appropriate model for frequencies between 1/1300 $d^{-1}$ and 1/75 $d^{-1}$ with noise strength $1.27 \times  10^{-21}$  Hz s$^{-2}$ ({\c{S}}ahiner et al. 2012).
For 4U 1907+09, {\c{S}}ahiner et al. (2012) suggested that at shorter time scales there may be a transient 
accretion disc formation around the neutron star which causes random walk
 in pulse frequency while the source has a  long term steady spin down trend. Her X-1 and 4U 1626-67 are disc accretors with low mass companions. For these sources it is found that pulse frequency time series are consistent with a random walk model with white noise strengths in the range $10^{-21}$ to $10^{-18}$ Hz s$^{-2}$, but red noise in pulse frequency derivative can not be excluded because of the sampling pulse frequencies and narrow range of power spectra. The disc accretor Cen X-3 has red noise in its pulse frequency derivatives and the noise strength varies from low to high
frequencies as $10^{-16}$ to $10^{-18}$ Hz s$^{-2}$ (Bildsten et al. 1997).
Power law index of power spectra of X Per is $\sim-1$. This implies that at short time scales disc accretion dominates and noise is less, on the other hand at
longer time scales greater than viscous time scales there is excessive noise. The steepest power law index for the power
spectra of pulse frequency derivatives among the HMXBs have been seen
 in SAX J 2103.5+4545 with power law index 2.13 (Baykal et
al. 2007). For GX 1+4 and OAO 1657-415 power law indices $\sim -1$ and
$\sim  0$, respectively. For these sources, accretion discs can be formed at short time scales (Baykal 1997, Bildsten et al., 1997). X Per has the lowest noise strength (or torque noise)   
among the HMXRBs discussed above (Baykal \& \"{O}gelman 1993, Bildsten et al., 1997). However the steep power law index or red noise in pulse frequency derivative 
suggests that it could have a transient accretion disc.
	
From the two average (summed) power spectra obtained from 411 day and 479 day long observations (see Figure \ref{ltpow}), we find that a power law model with two 
breaks that correspond to transitions from $f^0$ to $f^{-1.4}$ and from $f^{-1.4}$ to $f^{-2}$ parts of the power 
spectra fits well to both power spectra.  These break frequencies are shown to be insensitive to X-ray luminosity of the source which may indicate that accretion geometry is not affected significantly as the accretion rate changes. 

It is common to observe breaks in power spectra of accretion powered pulsars (see Revnivtsev et al. 2009 and references therein). Two breaks observed in the average power spectra of X Per may be an indication of the fact that the source has at least two different accretion flow components dominating the overall flow. These breaks sometimes also correspond to different spectral states of sources if power spectra are generated from data with a sufficiently long time interval (see \.{I}\c{c}dem et al. 2011). 

The pulse fractions show a clear correlation with source count rate. As the flux increases the material appears to accrete more efficiently on the neutron star's magnetic
poles.  At higher mass accretion rates, the accreting plasma should be interacting with the magnetosphere closer to the neutron star. The shorter distance and the stronger field there could contribute to a higher fraction flowing to the poles. 

\section*{Acknowledgment}

We acknowledge support from T\"{U}B\.{I}TAK, the Scientific and Technological Research Council of Turkey through the 
research project TBAG 109T748. We thank Tod Strohmayer and Craig Markwardt for useful suggestions.

\bsp

\label{lastpage}


\begin{thebibliography}{99}


\bibitem{} Alpar M. A., Shaham J., 1985, Nature, 317, 681
\bibitem{} Baykal A., \.{I}nam S. \c{C}., Stark M. J., Heffner C. M., Erkoca A. E., Swank J. H., 2007, MNRAS, 374, 1108
\bibitem{} Baykal A., \"{O}gelman H., 1993, A\&A, 267, 1
\bibitem{} Bildsten L., Chakrabarty D., Chiu J., Finger M.H., Koh D.T., Nelson R.B., Prince T.A, Rubin B.C, Scott, D.M., Stollberg M., Vaughan, B.A., Wilson C.A., Wilson R.B., 1997, ApJ, 113, 367
\bibitem{} Coburn W., Heindl W.A., Gruber D., Rothschild R.E., Staubert R., Wilms J., Kreykenbohm I., 2001, ApJ, 552, 738
\bibitem{} Cordes J. M., 1980, ApJ, 237, 216
\bibitem{} Deeter J. E., 1981, Pulse timing techniques and applications to X-ray pulsars. Ph.D. Thesis. Washington University: USA.
\bibitem{} Deeter J. E., Boynton P. E., 1982, ApJ, 261, 337
\bibitem{} Deeter J. E., 1984, ApJ, 281, 482
\bibitem{} Deeter J. E., Boynton P. E., 1985, in Hayakawa S. and Nagase F., Proc. Inuyama Workshop: Timing Studies of X-Ray
Sources, p.29, Nagoya Univ., Nagoya
\bibitem{} Deeter J. E., Boynton P. E., Lamb F. K., Zylstra G., 1989, ApJ, 336, 376
\bibitem{} Delgado-Marti H., Levine A.M., Pfahl E., Rappaport S.A., 2001, ApJ, 546, 455
\bibitem{} Di Salvo T., Burderi L., Robba N.R., Guainazzi M., 1998, ApJ, 509, 897
\bibitem{} Doroshenko V., Santangelo A., Kreykenbohm I., Doroshenko R., 2012, A\& AL, 540, 1 
\bibitem{} Esposito P., Israel G.L., Sidoli L., Mason E., Rodriguez Castillo, G.A., Halpern J.P., Moretti A., Gotz D., 2013, MNRAS, 433, 2028
\bibitem{} Ghosh P., Lamb F., 1979, ApJ, 234, 296
\bibitem{} Haberl F., 1994, A\& A, 283, 175
\bibitem{} \.{I}\c{c}dem B., Baykal A., 2011, A\&A, 529, A7
\bibitem{} \.{I}nam S.\c{C}., Baykal A., Swank J., Stark M.J., 2004, ApJ, 616, 463 
\bibitem{} Jahoda K., Swank J. H., Giles A. B., Stark M. J., Strohmayer T., Zhang W., Morgan E. H., 1996, Proc. SPIE, 2808, 59
\bibitem{} Jahoda K., Markwardt C.B., Radeva Y., Rots, A.H., Stark M.J., Swank J.H., Strohmayer T.E., Zhang W. 2006, ApJS, 163, 401
\bibitem{} Leahy D. A., Darbro W., Elsner R. F., Weisskopf M. C., Kahn S., Sutherland P. G., Grindlay J. E., 1983, ApJ, 266, 160
\bibitem{} Lutovinov A.A., Grebenev S.A., Sunyaev R.A., Pavlinsky M.N., 1994, Astron. Lett., 20, 538
\bibitem{} Lutovinov A., Tsygankov S., Chernyakova M., 2012, MNRAS, 423, 1978
\bibitem{} Nagase F., 1989, PASJ, 41, 1 
\bibitem{} Revnivtsev M., Churazov E., Postnov K., Tsygankov S., 2009, A\&A, 507, 1211
\bibitem{} Robba N.R., Burderi L., Wynn G.A. et al. 1996, ApJ, 472, 341
\bibitem{} {\c{S}}ahiner S., {\.{I}}nam S.{\c{C}}., Baykal A., 2012, MNRAS, 421, 2079
\bibitem{} Shakura N., Postnov K., Kochetkova A., Hjalmarsdotter L., 2012, MNRAS, 420, 216
\bibitem{} Takeshima T., 1998. The Hot Universe. In Katsuji K., Shunji K., Masayuki I. (Ed), Proceedings of the IAU Symposium 188 (pp.368). Dordrecht: Kluwer Academic.
\bibitem{} Ubertini P., Lebrun F., Di Cocco G. et al., 2003, A\&A, 411, L131
\bibitem{} van der Klis, M., Stella, L., White, N., Jansen, F., Parmar, A. N. 1987, ApJ,
316, 411
\bibitem{} van der Klis, M. 1989, in Timing Neutron Stars, ed. H. \"{O}gelman \& E. P. J.
van den Heuvel (Dordrecht: Kluwer), 27
\bibitem{} White N.E., Mason K.E., Sanford P.W., Murdin P., 1976, MNRAS, 176, 20

\end{thebibliography}
\end{document}